\documentclass[12pt]{revtex4-1}
\usepackage{amssymb,graphicx,xspace,color,amsmath}

\usepackage{natbib}
\usepackage{lineno}

\newcommand{\tc}{$T_c$\xspace}
\newcommand{\tn}{$T_N$\xspace}
\newcommand{\ts}{$T_S$\xspace}
\newcommand{\tnem}{$T_{nem}$\xspace}
\newcommand{\ba}{BaFe$_2$As$_2$\xspace}
\newcommand{\barh}{Ba(Fe$_{1-x}$Rh$_x$)$_2$As$_2$\xspace}

\newcommand{\la}{LaFeAsO\xspace}
\newcommand{\laco}{LaFe$_{1-x}$Co$_x$AsO\xspace}

\begin{document}


\textbf{\LARGE{Ubiquitous enhancement of nematic fluctuations across the phase diagram of iron based superconductors probed by the Nernst effect}}\\
\\
\\Christoph Wuttke$^{1}$, Federico Caglieris$^{1,2}$, Steffen Sykora$^{1,3}$, Frank Steckel$^{1}$, Xiaochen Hong$^{1,4}$, Sheng Ran$^{5,6}$, Seunghyun Khim$^{1,7}$, Rhea Kappenberger$^{1}$, Sergey L. Bud'ko$^{5}$, Paul C. Canfield$^{5}$, Sabine Wurmehl$^{1}$, Saicharan Aswartham$^{1}$, Bernd B\"{u}chner$^{1,8,9}$ \& Christian Hess$^{1,4,8}$\\
\\
$^{1}$\textit{Leibniz-Institute for Solid State and Materials Research, IFW-Dresden, 01069 Dresden, Germany}\\
$^{2}$\textit{CNR-SPIN, 16152 Genova, Italy}\\
$^{3}$\textit{Institut f\"ur Theoretische Physik, TU Dresden, 01069 Dresden, Germany}\\
$^{4}$\textit{Fakult\"at f\"ur Mathematik und Naturwissenschaften, Bergische Universit\"at Wuppertal, 42097 Wuppertal, Germany}\\
$^{5}$\textit{Ames Laboratory and Department of Physics and Astronomy, Ames, Iowa, 50011, USA}\\
$^{6}$\textit{Department of Physics, Washington University, St. Louis, Missouri, 63130, USA}\\
$^{7}$\textit{Max Planck Institute for Chemical Physics of Solids, 01187 Dresden, Germany}\\
$^{8}$\textit{Center for Transport and Devices, TU Dresden, 01069 Dresden, Germany}\\
$^{9}$\textit{Institut f\"{u}r Festk\"{o}rperphysik, TU Dresden, 01069 Dresden, Germany}\\
\\




\textbf{The role of nematic fluctuations for unconventional superconductivity has been subject of intense discussions for many years. In iron-based superconductors, the most established probe for electronic-nematic fluctuations, i.e. the elastoresistivity seems to imply that superconductivity is reinforced by electronic-nematic fluctuations, since the elastoresistivity amplitude peaks at or close to optimal \tc. However, on the over-doped side of the superconducting dome, the diminishing elastoresistivity suggests a negligible importance in the mechanism of superconductivity. Here we introduce the Nernst coefficient as a genuine probe for electronic nematic fluctuations, and we show that the amplitude of the Nernst coefficient tracks the superconducting dome of two prototype families of iron-based superconductors, namely Rh-doped \ba and Co-doped \la. Our data thus provide fresh evidence that in these systems nematic fluctuations foster the superconductivity throughout the phase diagram.}\\

Unravelling the interaction of electronic orders in the phase diagram of copper-based~\cite{Daou2010, Lawler2010} and iron-based materials~\cite{Chuang2010}, such as stripe order~\cite{Tranquada1995} or nematicity~\cite{Fernandes2014, Orth2019, Kivelson1998} on the one hand and superconductivity on the other hand, became one of the most important tasks in understanding high-temperature superconductors. Most families of iron-based superconductors show a nematic phase in close vicinity to the superconducting transition, with nematic fluctuations being present in large areas of their electronic phase diagram. Therefore, nematic fluctuations have been widely proposed to be coupled to superconductivity and to be an essential ingredient to the pairing of electrons in these materials~\cite{Lederer2015, Yamase2013, Dumitrescu2016, Fernandes2013a}. However, experimental evidence for nematic fluctuations in the overdoped superconducting regime is scarce and mostly limited to the underdoped regions. It remains thus elusive whether such fluctuations are an indispensable ingredient in the Cooper pairing or play the role of additionally enhancing the superconductivity, potentially driven by antiferromagnetic fluctuations.  

\begin{figure}
\includegraphics[width=1\textwidth]{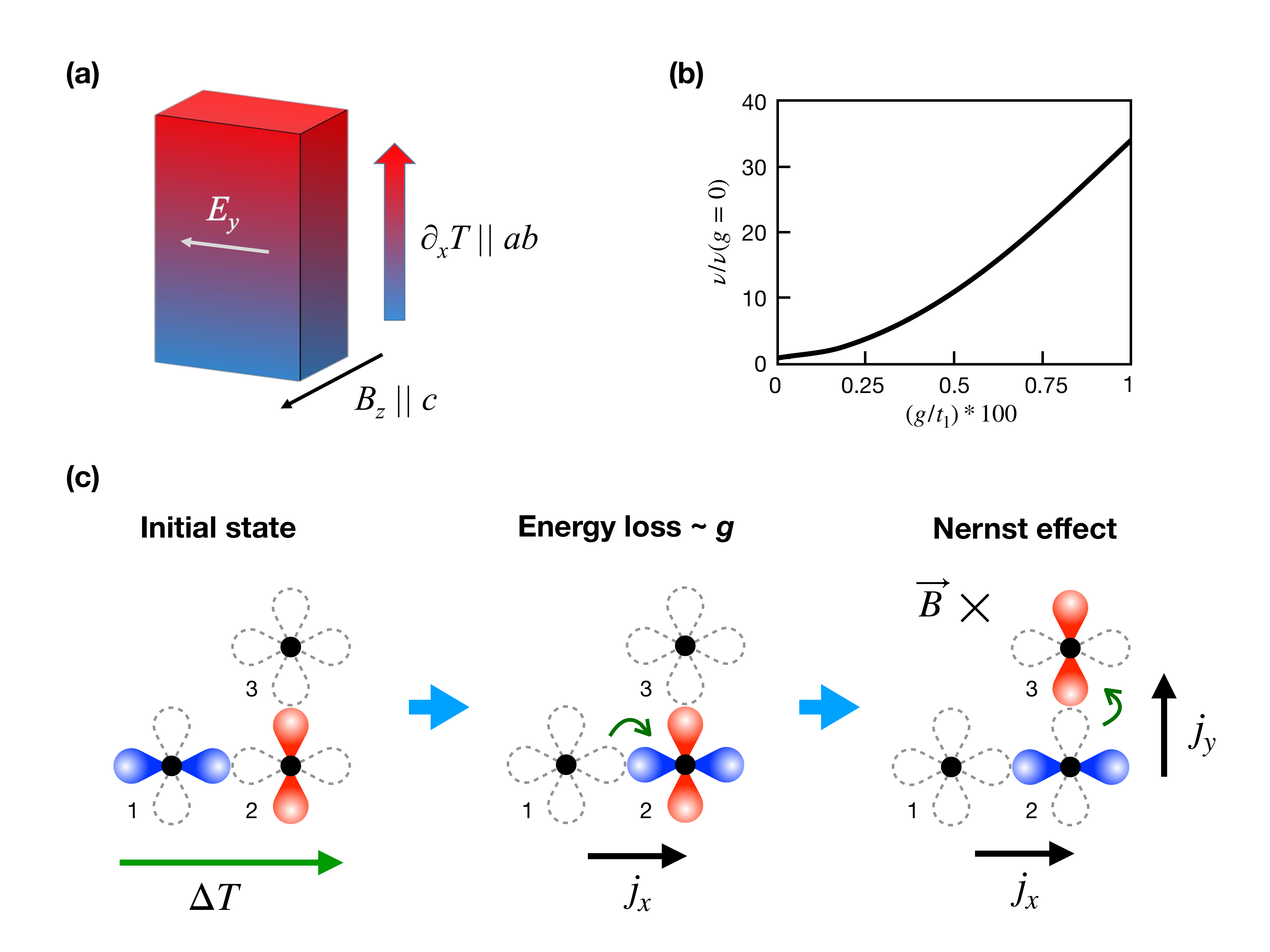}
\caption{\textbf{(a)} Schematic Nernst setup for measuring the Nernst coefficient $\nu$. A resistive chip heater is used to heat up the upper side of the sample (red). The bottom is coupled to a thermal bath of controlled temperature (blue), therefore a temperature gradient arises in the $ab$-plane. The magnetic field is applied in $c$-direction and the Nernst signal is measured perpendicular to the temperature gradient in the $ab$-plane. \textbf{(b)} Calculated Nernst coefficient $\nu$ as a function of the strength $g$ of the nematic fluctuations. Starting from very small values $\nu$ increases dramatically. \textbf{(c)} Schematic picture of the mechanism leading to the effect in panel (b). An external temperature gradient $\Delta T$ generates a preferred hopping of conduction electrons (left panel). If the neighbouring lattice site is already  occupied, an intermediate state with double occupancy is created (middle panel). Due to the Lorentz force of the perpendicularly oriented magnetic field $\vec{B}$ the enhanced energy from nematic fluctuations is preferably reduced by removing the electron in the perpendicular orbital (right panel) leading to enhanced transverse transport.}
\label{fig:Concept}
\end{figure}

The Nernst effect describes the occurrence of a transverse electric field $E_{y}$ perpendicular to a temperature gradient $\left|\partial_{x} T\right|$ and perpendicular to an external magnetic field $B_z$ with the Nernst signal $N$ given by $N = E_y / |\partial_{x}T|$ (see Fig.~\ref{fig:Concept}(a)). In the linear regime with respect to  $B_z$ one introduces the Nernst coefficient $\nu$ as $\nu = N / B_{z}$. The Nernst effect is therefore a transverse transport probe which combines thermal and charge excitations and is strongly influenced by electronic states near the Fermi level. Hence, it is expected that the Nernst effect is sensitive to fluctuations of electronic order parameters and of the Fermi surface~\cite{Hackl2009a, Hackl2009b, Hackl2010,Behnia2016,Xu2000}. Indeed, in pioneering experiments the Nernst effect has successfully been employed for probing stripe/nematic phases and pertinent Fermi surface reconstructions in cuprate superconductors~\cite{Cyr-Choiniere2009,Daou2010,Matusiak2009}. We will show below that the Nernst effect allows probing nematic fluctuations in iron-based superconductors and that it provides profound information on the interaction of nematicity and superconductivity.  

To obtain a conceptional insight on the influence of nematic fluctuations on the Nernst coefficient $\nu$ in iron-based superconductors, we consider a model for nematic fluctuations, where conduction electrons can occupy the $d_{xz}$ and $d_{yz}$ orbitals of the iron atoms on a square lattice. Such a minimal model captures basic features of the band structure of iron-based superconductors and allows for a ground state with nematic order~\cite{Raghu2008,Dumitrescu2016}. The model Hamiltonian reads~\cite{Dumitrescu2016},
\begin{eqnarray}
\label{Model_H}
\mathcal{H} &=& -\sum_{{\bf i}{\bf j},ab,\sigma} \left( t_{{\bf i}{\bf j}}^{ab} c_{{\bf i}a\sigma}^\dag c_{{\bf j}b\sigma}^{} + \mbox{h.c.} \right) - \mu \sum_{{\bf i},a} n_{{\bf i},a} - \frac{g}{2} \sum_{\bf i} \left(n_{{\bf i},xz} - n_{{\bf i},yz} \right)^2 = \mathcal{H}_0 + \mathcal{H}_g,
\end{eqnarray}
where $a,b = xz,yz$ denote the orbital indices, $\sigma = \uparrow , \downarrow$ the spin index, and $n_{{\bf i},a} = \sum_{\sigma} c_{{\bf i}a\sigma}^\dag c_{{\bf i}a\sigma}^{}$ is the occupation of orbital $a$ on lattice site ${\bf i}$. The first term in Eq.~\eqref{Model_H} describes the kinetic energy of the conduction electrons. In order to start from a situation which is relevant to a variety of iron-based superconductors we take the hopping parameters $t_{{\bf i}{\bf j}}^{ab}$ from Ref.\onlinecite{Yamase2013}, leading to the usual band structure of the iron-pnictides consisting of one hole pocket around the $\Gamma$ point and electron pockets around $X$- and $Y$-points of the Brillouin zone. The chemical potential $\mu$ of the system is included in the second term. A variation of $\mu$ changes the relative size of the hole and electron pockets in momentum space, and therefore, the electron filling. A variation of $\mu$ can be achieved experimentally by doping or applying pressure. 

Most importantly, the third term in Eq.~\eqref{Model_H} accounts for the nematic fluctuations. Due to its quadratic form and negative sign, this term energetically favors (proportional to the coupling strength $g$) a difference in the local occupation between the $d_{xz}$ and $d_{yz}$ orbitals. Thus, if one orbital at lattice site ${\bf i}$ is preferentially occupied, occupation of the orbital in perpendicular direction is unfavoured. Together with the hopping term, such an effective electron-electron interaction captures the basic property of nematicity. We emphasize at this point that despite the local character of the nematic interaction our theoretical treatment of the model goes beyond single site fluctuations. Due to the presence of the itinerant hopping term which combines not only different lattice sites but also different orbitals our analysis naturally takes into account non-local higher order processes. Since the rotational symmetry is not broken by the Hamiltonian \eqref{Model_H}, the introduced model can be thought of being relevant for an iron-pnictide material which is close to a nematic instability. 

Using this model we find (see Supplementary information for details) that the Nernst coefficient $\nu$ is sensitive to the occupation difference of the orbitals and is strongly enhanced by a finite nematic coupling $g$, which is shown in Fig.~\ref{fig:Concept}(b). Taken into account the typical energy scales of phonons (10 meV) and the nearest-neighbour hopping matrix element (1 eV), $g/t_1$ can be expected to be in the order of 0.01. This enhancement of $\nu$ can be qualitatively explained by the process displayed in Fig.~\ref{fig:Concept}(c). Due to the nematic coupling electrons that are moving along a temperature gradient and are occupying two different orbitals at the same lattice site will demand an additional energy cost proportional to $g$, which is avoided by a further movement of one electron to a neighbouring lattice site. The presence of a perpendicular magnetic field $\vec{B} || z$ leads to a coupling between $xz$ and $yz$ orbitals and therefore to an enhanced hopping perpendicular to the temperature gradient as well as the magnetic field. 

Inspired by the above considerations, we performed Nernst effect measurements (experimental setup schematically shown in Fig.~\ref{fig:Concept}(a)) in all important regions of the electron doped phase diagram on \barh single crystals. Rh-doped \ba represents a canonical electron-doped iron-based superconductor with a phase diagram and superconducting properties which virtually are identical to those of Co-doped \ba~\cite{Ni2009, Kim2018}. In particular, the spin density wave (SDW) ground state of the parent compound which sets in at the magnetic/structural transitions at $T_N\approx T_S\approx 135~$K is gradually suppressed upon Rh-doping, in favour of a superconducting state at $x\gtrsim 2.5$~\%. 

\begin{figure}
\includegraphics[width=\textwidth]{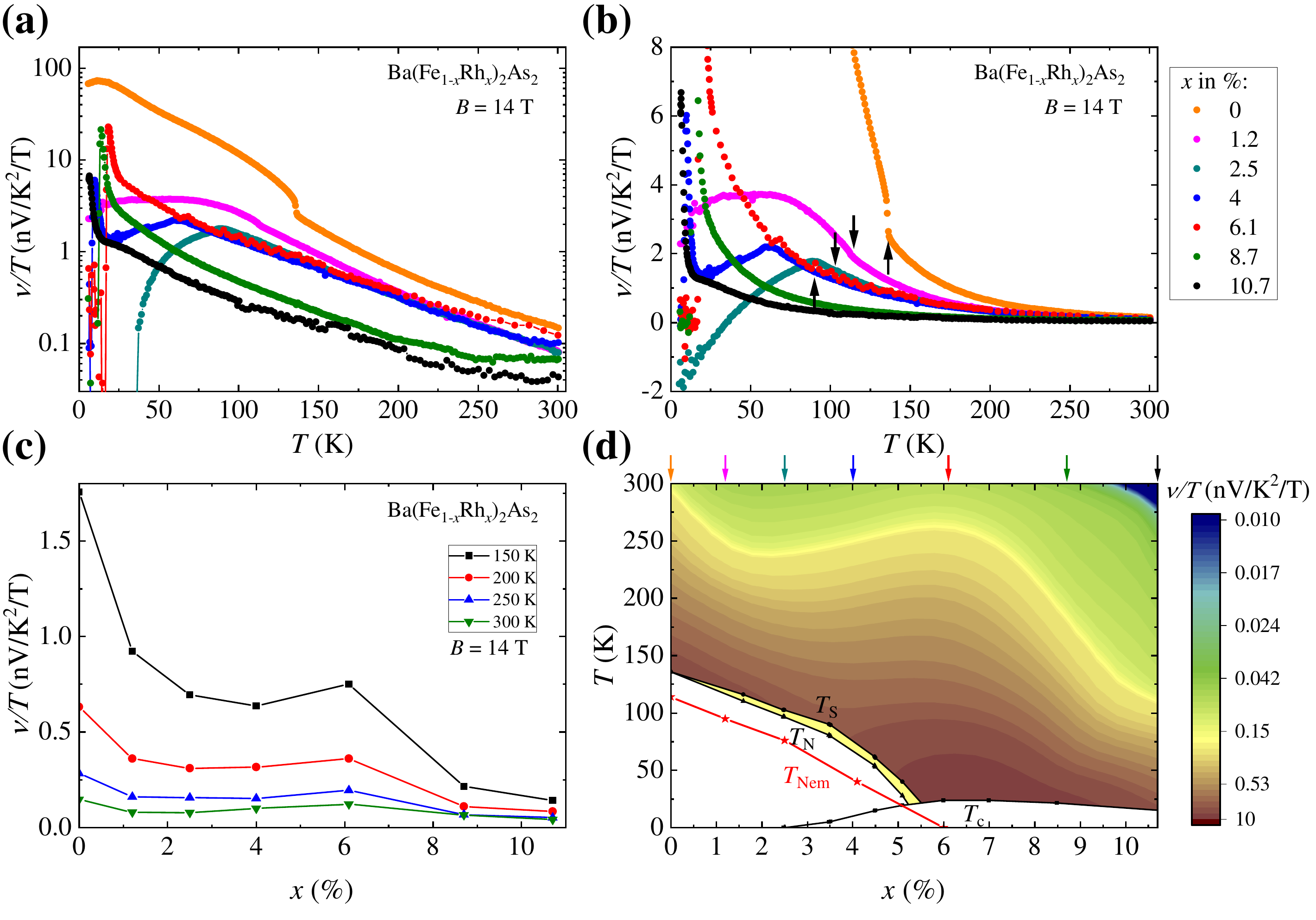}
\caption{Nernst coefficient divided by temperature $\nu/T$ of \barh as a function of $T$ and $x$. All data measured along the $ab$-plane with a magnetic field of $B=14~\mathrm{T}$ applied parallel to the $c$-axis. \textbf{(a)} $\nu/T$ as a function of temperature. \textbf{(b)} Zoom-in on the data shown in panel (a). Black arrows indicate structural phase transitions. \textbf{(c)} $\nu/T$ vs. Rh doping level $x$ at selected temperatures. \textbf{(d)} Colour coded phase diagram of the Nernst coefficient divided by temperature $\nu/T$ as a function of Rh doping level $x$ and temperature $T$. Coloured arrows on the top mark show the nominal doping of the measured samples. \ts, \tn and \tc data taken from Ref.~\onlinecite{Ni2009}. \tnem data is obtained from elastoresistivity measurements on the same samples (compare supplemental information).}
\label{fig:Ba122_Nu_data}
\end{figure}

To take into account the effect of entropy which leads to an intrinsic linear temperature dependence, Fig.~\ref{fig:Ba122_Nu_data}(a) displays the temperature dependence of the Nernst coefficient divided by temperature $\nu/T$ for different Rh-doping levels in \barh on a semi-logarithmic scale. At most instances $\nu/T$ is large and positive, and steadily increases by 1-2 orders of magnitude upon cooling. Further inspection reveals that this steady increase is always present whenever a sample is in the tetragonal paramagnetic, non-superconducting phase, i.e. at $T>T_S$ for $x\leq 4$~\% and $T>T_c$ for $x > 4$~\% (see Fig.~\ref{fig:Ba122_Nu_data}(b)). In contrast, in the magnetically ordered orthorhombic phase either large positive or negative contributions occur in the Nernst effect. In the case of nematic order, an interpretation of these low-temperature contributions is elusive since a strong in-plane anisotropy, uncontrolled twinning and possible additional contributions from highly mobile Dirac-like fermions need to be considered~\cite{Matusiak2019a, Matusiak2019b, Caglieris2021}. On the other hand, the strong additional increase which is well recognizable near $T_c$ of the superconducting samples is well-known to result from vortex motion and therefore is of no further interest in the present study. While the Nernst effect is in principle sensitive to fluctuating Cooper pairs above the critical temperature and is capable of tracking the transition from vortex liquid to a phase-fluctuating superconducting regime~\cite{Behnia2016, Xu2000}, it is rather the strong enhancement of the Nernst coefficient at high temperatures, far above any superconducting fluctuation regime, which attracts our attention. This enhancement is strongest in the parent compound of both families, where the ground state shows long range antiferromagnetic order rather than superconductivity. In view of our theory and the known presence of nematic fluctuations we expect the amplitude of these fluctuations to be directly encoded in the Nernst signal. 

Thus motivated by this, we have investigated the doping evolution of several isotherms of $\nu/T$ as shown in Fig.~\ref{fig:Ba122_Nu_data}(c). The data reveal a strong initial reduction of $\nu/T$ at low doping levels which tends to saturate upon reaching superconducting doping levels ($x\leq 4$~\%). However, a significantly larger $\nu/T$ occurs at $x=6.1$~\% before being strongly suppressed at $x>8$~\%. This non-monotonic doping dependence is especially enhanced at high temperatures ($T\gtrsim 200~\mathrm{K}$). Remarkably, as seen in Fig.~\ref{fig:Ba122_Nu_data}(d) (please note the colour coding of the amplitude for better visibility), $\nu/T$ tracks the superconducting dome and shows a maximum near the optimal doping level $x\approx 6$~\%. These experimental findings therefore suggest a relationship between $\nu/T$ and the superconducting transition $T_c$, whereas our theoretical finding points out the sensitivity of $\nu/T$ to nematic fluctuations. In a nutshell, this comparison suggests that electronic-nematic fluctuations play an important role for enhancing superconductivity. 

To verify the universal character of our result across different families of iron-based superconductors, we performed Nernst measurements on high-quality single crystals of Co-doped \la ~\cite{Kappenberger2018}. \la is one of the most famous representatives of iron based systems, since superconductivity of this material class was first discovered in F-doped \la~\cite{Kamihara2008}. Another way of inducing a superconducting ground state in \la is replacing Fe with Co~\cite{Sefat2008}. In contrast to \ba, \la is showing separated magnetic and structural transitions. Both are gradually suppressed as a function of Co content $x$ and vanish in a region $x\approx 4.5$~\% - $5$~\%, a superconducting dome develops above $x\approx 5$~\% with an optimal doping level of $x = 6$~\%~\cite{Kappenberger2018, Hong2020, Scaravaggi2021, Lepucki2021}.  

\begin{figure}
\includegraphics[width=\textwidth]{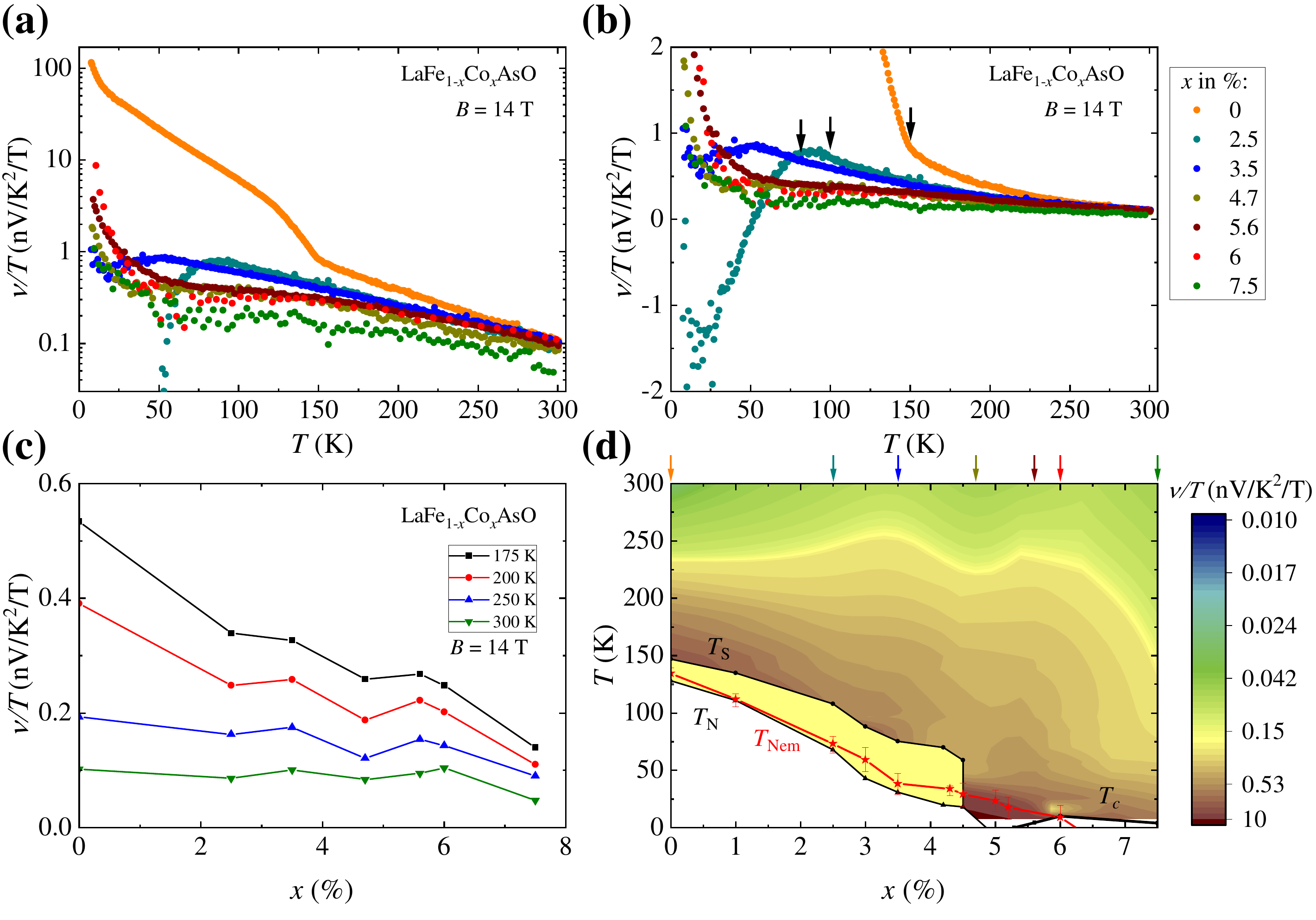}
\caption{Nernst coefficient divided by temperature $\nu/T$ of \laco as a function of $T$ and $x$. All data measured along the $ab$-plane with a magnetic field of $B=14~\mathrm{T}$ applied parallel to the $c$-axis. \textbf{(a)} $\nu/T$ as a function of temperature on a normal and logarithmic scale (inset). \textbf{(b)} Zoom-in on the data shown in panel (a). Black arrows indicate structural phase transitions. \textbf{(c)} $\nu/T$ vs. Co doping level $x$ at selected temperatures. \textbf{(d)} Colour coded phase diagram of the Nernst coefficient divided by temperature $\nu/T$ as a function of Co doping level $x$ and temperature $T$. Coloured arrows on the top mark show the nominal doping of the measured samples. \ts, \tn, \tc and \tnem data taken from Ref.~\onlinecite{Hong2020}.}
\label{fig:La1111_Nu_data}
\end{figure}

In analogy to the 122 system, two temperature regimes are excluded from the analysis: (i) due to the formation of twinned domains and a strong in-plane anisotropy the data in the low temperature nematic phase is elusive and (ii) $\nu/T$ in the superconducting phase as well as close to \tc is governed by well-known contributions due to vortex motion and superconducting fluctuations~\cite{Behnia2016}. Hence, we focus on the steady increase of $\nu/T$ by more than one order of magnitude upon lowering the temperature in the tetragonal phase. Fig.~\ref{fig:La1111_Nu_data}(a) shows data of $\nu/T$ vs. $T$ for all \laco compositions on a semi-logarithmic scale. Similar to \barh, $\nu/T$ displays a large positive value for all doping levels at room temperature, which increases monotonically upon cooling as long as a sample is in the tetragonal paramagnetic, non-superconducting phase. The temperature dependencies of $\nu/T$ in \barh and \laco are showing strong similarities as well. Apart from a slight difference in magnitude, the features obtained by the temperature dependent measurements seem to be an universal property of the Nernst effect in iron based superconductors. In fact, similar levels of electron doping (Rh doping in \ba and Co doping \la) produce remarkably similar $\nu/T$ vs. $T$ curves (for example in both systems the doping level $x = 2.5$~\% is the only composition where a negative Nernst signal is obtained). The extremely similar behaviour in \barh and \laco, i.e. the evolution of $\nu/T$ as function of temperature and doping, seems to be of universal character and is therefore strongly suggesting that the same mechanism is driving the large Nernst coefficient due to the presence of nematic fluctuations.   

Furthermore, the doping dependent data of \laco (Fig.~\ref{fig:La1111_Nu_data}(c) and (d)) reveal a non-monotonic doping dependence as well, albeit some differences. On one hand, the variation as a function of doping is smaller compared to \barh, though an enhancement of $\nu/T$ above the superconducting dome is also visible in \laco, which suggests that a correlation between the maximum of $\nu/T$ and \tc is present as well. On the other hand, the data for \laco even seem to indicate a slight double peak behaviour as a function of doping, which resembles recent reports of elastoresistivity, i.e., an alternative well-established probe for nematic fluctuations~\cite{Chu2010, Chu2012, Kuo2016, Hong2020}, in \laco (measured on a superset of the crystals that were used in this work)~\cite{Hong2020}.

The existence of the same double peak structure in two distinct probes suggest that this is a robust feature of the nematic fluctuations in this system. Motivated by the analogy between Nernst and elastoresistance in \laco, we checked the consistency of the two techniques in \barh. Interestingly we observed a mismatch, namely the Nernst coefficient peaks at the optimal doping, as mentioned above, while $\chi_{\rho}$ peaks in the underdoped region and does not track the superconducting dome, in agreement with previous reports on Co-doped \ba~\cite{Chu2010, Chu2012}.   

Clearly, this observation implies that elastoresistivity and Nernst effect probe similar, yet distinct aspects of the nematic fluctuations, where apparent differences are likely related to subtleties of the considered systems. This conclusion is corroborated by recent elasto-Seebeck and elasto-Nernst data which possess a somewhat different coupling to the nematic susceptibility than the elastoresistivity~\cite{Caglieris2021}. In this context it is important to consider that all the elasto-transport properties, including elastoresistivity, crucially depend on the electron-lattice coupling, since, by definition, they represent the response of an electronic system to an external uniaxial strain. This is not the case for the pure Nernst effect, which is measured without stressing the lattice. In this sense, the Nernst coefficient can be sensitive to nematic fluctuations even far away from the actual structural transition, where the electron-lattice coupling is weaker.

We further emphasize that, on a more general perspective, both scattering time and Fermi surface distortions and fluctuations thereof have to be taken into account to explain the transport nematic phenomenology. Since thermoelectric properties, especially the Nernst effect, are known to be extremely sensitive to the latter~\cite{Hackl2009a, Hackl2009b} it is natural to consider an orbital contribution, which eventually is connected to Fermi surface distortions, as a first candidate to theoretically approach the influence of nematicity on the Nernst coefficient in our basic model. We point out, as is discussed in Ref.~\onlinecite{Caglieris2021}, that an anisotropic scattering time can be expected to play a further important role in these systems. 

We mention further, that, beside a finite nematic coupling, another important ingredient in determining the electronic properties of iron based superconductors is their multi-band nature. From the orbital point of view, our minimal model including a finite nematic coupling, arises within a two-band system, therefore it is intrinsically a multi-band effect. However, one could also argue that the evolution of Nernst coefficient as a function of doping could arise either from a doping-dependent charge-carrier compensation or a doping-dependent scattering rate. It is well-established that the ambipolar flow of electron-like and hole-like quasiparticles enhances the Nernst coefficient, unlike the Seebeck and the Hall effect, where the contributions of carriers with different signs tend to cancel out. This peculiar property of the Nernst coefficient makes its doping dependence predictable, namely a progressive departure from the multi-band compensation due to doping should lead to a monotonic decrease of the effect. In our case, this scenario seems to be insufficient, since we observed a pronounced non-monotonic doping-dependence of the Nernst coefficient. We could now suppose that the doping dependence of the Nernst coefficient could be caused by a change of the scattering rate with the Rh- or the Co-content. However, within our model, we should introduce a non-monotonic change of the scattering time up to one order of magnitude to produce an effect comparable to the influence of a finite nematic coupling (see Figure~S5 in the Supplemental material). Even admitting a different scattering rates for the electron- and hole-like bands, such a huge change of the scattering time would be unusual and incompatible with the monotonic doping dependence of the resistivity of electron-doped iron-based superconductors \cite{Aswartham2011}. Hence, although one cannot exclude a priori that a doping-dependence of the charge carrier compensation and of the scattering time can play a role, a finite nematic coupling seems to be essential to reproduce our data.

Finally we mention that antiferromagnetic fluctuations in principle can give rise to a Nernst signal, too~\cite{Hackl2009a, Daou2015, Steckel2014}. However, it is well established by nuclear magnetic resonance (NMR) that in electron doped Ba122 and La1111 the low-energy spin-fluctuations are \textit{monotonically} suppressed towards optimal doping~\cite{Dioguardi2015, Lepucki2021}. In contrast to this, our Nernst effect data clearly exhibit a \textit{non-monotonic} doping dependence near optimal doping. Furthermore, the measured Nernst coefficient in the tetragonal phase always is linear in magnetic field, in contrast to reports where a non-linear in field Nernst response is interpreted as evidence of an enhancement due to spin fluctuations~\cite{Daou2015}. Thus, while we cannot exclude spin fluctuations to contribute to the enhanced Nernst effect at low doping levels, such can be ruled as the origin for the enhanced Nernst effect near optimal doping.

In conclusion, we established the Nernst effect as a sensitive probe for nematic fluctuations, and we we show that the amplitude of the Nernst coefficient in the normal state tracks the superconducting dome in the electron doped phase diagram of two representatives of major families of the iron based superconductors. Our results therefore imply that nematic fluctuations are an indispensable ingredient for enhancing \tc in the iron based superconductors. Furthermore, our analysis suggests the Nernst effect as a principal technique to probe nematicity and thus it should be considered for future experiments in order to unravel the mysteries of electronic order in iron based superconductors.


%

\newpage

\textbf{Acknowledgments} We acknowledge discussions with P. Carretta as well as technical support by D. Baumann and T. Schreiner. This work has been supported by the Deutsche Forschungsgemeinschaft (DFG) through the DFG Research Grant CA1931/1-1 (F.C.) as well as through project C07 of SFB 1143 (project ID 247310070) and priority program DFG- GRK1621. This project has received funding from the European Research Council (ERC) under the European Unions’ Horizon 2020 research and innovation program (grant agreement No 647276-MARS-ERC-2014-CoG). S.S. acknowledges financial support by the Deutsche Forschungsgemeinschaft via the Emmy Noether Programme ME4844/1-1 (project id 327807255). S.A. acknowledges (DFG) through Grant No.AS 523/4-1. Work done at Ames Laboratory (P.C.C., S.L.B., S.R.) was supported by the U.S. Department of Energy, Office of Basic Energy Science, Division of Materials Sciences and Engineering.  Ames Laboratory is operated for the U.S. Department of Energy by Iowa State University under Contract No. DE-AC02-07CH11358.\\

\textbf{Author contributions} C.W., F.C., F.S. and X.C.H. performed the measurements, S.S. developed the theoretical model. S.R., S.K., R.K., S.L.B., P.C.C., S.W. and S.A. grew and characterized the samples. B.B. and C.H. designed the project, C.W., F.C., F.S., S.S. and C.H. wrote the manuscript.\\

\textbf{Competing Interests} The authors declare that they have no competing financial or non-financial interests.\\

\textbf{Data Availability} The data that support the findings of this study are available from the corresponding author upon reasonable request.\\

\textbf{Correspondence} Correspondence should be addressed to C. Wuttke~(email: c.wuttke@ifw-dresden.de) or C. Hess~(email: c.hess@uni-wuppertal.de).\\


\end{document}



\textbf{\LARGE{Ubiquitous enhancement of nematic fluctuations across the phase diagram of iron based superconductors probed by the Nernst effect - Supplemental Information}}\\
\\
\\Christoph Wuttke$^{1}$, Federico Caglieris$^{1,2}$, Steffen Sykora$^{1,3}$, Frank Steckel$^{1}$, Xiaochen Hong$^{1,4}$, Sheng Ran$^{5,6}$, Seunghyun Khim$^{1,7}$, Rhea Kappenberger$^{1}$, Sergey L. Bud'ko$^{5}$, Paul C. Canfield$^{5}$, Sabine Wurmehl$^{1}$, Saicharan Aswartham$^{1}$, Bernd B\"{u}chner$^{1,8,9}$ \& Christian Hess$^{1,4,8}$\\
\\
$^{1}$\textit{Leibniz-Institute for Solid State and Materials Research, IFW-Dresden, 01069 Dresden, Germany}\\
$^{2}$\textit{CNR-SPIN, 16152 Genova, Italy}\\
$^{3}$\textit{Institut f\"ur Theoretische Physik, TU Dresden, 01069 Dresden, Germany}\\
$^{4}$\textit{Fakult\"at f\"ur Mathematik und Naturwissenschaften, Bergische Universit\"at Wuppertal, 42097 Wuppertal, Germany}\\
$^{5}$\textit{Ames Laboratory and Department of Physics and Astronomy, Ames, Iowa, 50011, USA}\\
$^{6}$\textit{Department of Physics, Washington University, St. Louis, Missouri, 63130, USA}\\
$^{7}$\textit{Max Planck Institute for Chemical Physics of Solids, 01187 Dresden, Germany}\\
$^{8}$\textit{Center for Transport and Devices, TU Dresden, 01069 Dresden, Germany}\\
$^{9}$\textit{Institut f\"{u}r Festk\"{o}rperphysik, TU Dresden, 01069 Dresden, Germany}\\
\\

\section{Theory}
\subsection{Minimal model}

We consider a two-band tight binding model, where conduction electrons can occupy the $d_{xz}$ and $d_{yz}$ orbitals of the iron atoms on a square lattice. Such a minimal model captures basic features of the band structure of iron-based superconductors and allows for a ground state with nematic order~\cite{Raghu2008,Dumitrescu2016}. The model Hamiltonian reads~\cite{Dumitrescu2016},
\begin{eqnarray}
\label{Model_H}
\mathcal{H} &=& -\sum_{{\bf i}{\bf j},ab,\sigma} \left( t_{{\bf i}{\bf j}}^{ab} c_{{\bf i}a\sigma}^\dag c_{{\bf j}b\sigma}^{} + \mbox{h.c.} \right) - \mu \sum_{{\bf i},a} n_{{\bf i},a} - \frac{g}{2} \sum_{\bf i} \left(n_{{\bf i},xz} - n_{{\bf i},yz} \right)^2 = \mathcal{H}_0 + \mathcal{H}_g,
\end{eqnarray}
where $a,b = xz,yz$ denote the orbital indices, $\sigma = \uparrow , \downarrow$ the spin index, and $n_{{\bf i},a} = \sum_{\sigma} c_{{\bf i}a\sigma}^\dag c_{{\bf i}a\sigma}^{}$ is the occupation of orbital $a$ on lattice site ${\bf i}$. The first term in Eq.~\eqref{Model_H} describes the kinetic energy of the conduction electrons. In order to start from a situation which is relevant to a variety of iron-based superconductors we take the hopping parameters $t_{{\bf i}{\bf j}}^{ab}$ from Ref.~\onlinecite{Yamase2013} leading to the usual band structure of the iron-pnictides consisting of one hole pocket around the $\Gamma$ point and electron pockets around $X$- and $Y$-points of the Brillouin zone. The chemical potential $\mu$ of the system is included in the second term. A variation of $\mu$ changes the relative size of the hole and electron pockets in momentum space and therefore also the electron filling. A variation of $\mu$ can be achieved experimentally for example by doping or applying pressure. 

Most importantly, the third term in Eq.~\eqref{Model_H} accounts for the nematic fluctuations. Due to its quadratic form and negative sign this term energetically favors (proportional to the coupling strength $g$) a difference in the local occupation between the $d_{xz}$ and $d_{yz}$ orbitals. Thus, if one orbital is particularly preferred at lattice site ${\bf i}$ the orbital referring to the perpendicular direction is at the same time avoided. Together with the hopping term such an effective electron-electron interaction captures the basic property of nematicity. Since the rotational symmetry is not broken by the Hamiltonian \eqref{Model_H} the introduced model can be thought of being relevant for an iron-pnictide material which is close to a nematic instability. 

The model \eqref{Model_H} has been studied theoretically by random phase approximation (RPA)~\cite{Yamase2013} and determinant quantum Monte Carlo (DQMC)~\cite{Dumitrescu2016}. As is well known the RPA works well in a weak-coupling regime. It predicts a variety of orders for Eq.~\eqref{Model_H}, depending on the value of $\mu$. In the range between $\mu / t_1 = 0.2$ and   $\mu / t_1 = 2.5$, where $t_1$ is the nearest neighbor intra-orbital hopping matrix element, the RPA predicts the onset of uniform nematic order for $g_c /t_1 \approx 1.7$. For $\mu / t_1 > 2.5$, susceptibility peaks at wave vectors $(0,\pi)$ and $(\pi,0)$ indicating stripe order have been found, while for $\mu / t_1 < 0.2$ the susceptibility peaks appear at wave vector $(\pi,\pi)$ and therefore predict antiferro-orbital (AFO) (antiferroquadrupole) order.

\subsection{Evaluation of the Nernst effect}

The Nernst coefficient $\nu = E_y / (-\nabla_x T)$ which is measured in this work can be generally expressed in terms of the thermoelectric tensor $\alpha_{ab}$ and charge conductivity tensor $\sigma_{ab}$ as
\begin{eqnarray}
\label{nu_general}
\nu &=& \frac{\alpha_{xy} \sigma_{xx} - \alpha_{xx} \sigma_{xy}}{\sigma_{xx}^2 + \sigma_{xy}^2}.
\end{eqnarray}
In the following, we consider a weak coupling regime and show that already a small value of $g$ (in comparison to the parameters of the kinetic energy) is sufficient to strongly enhance the Nernst coefficient $\nu$. As discussed above, for such a small coupling regime the properties of the model Hamiltonian \eqref{Model_H} are well-captured by the RPA which is a form of a mean field theory. Motivated by this result we proceed  to study the influence of nematic fluctuations on the Nernst coefficient \eqref{nu_general} as follows: At first we apply the mean-field approximation to decouple the interaction term in \eqref{Model_H} and to solve the model. Based on this decoupling we calculate the transport coefficients in Eq.~\eqref{nu_general} separately. Thereby we consider a relatively small magnetic field $B$ that produces the Nernst effect such that  perturbation theory up to lowest order in $B$ is sufficient to describe the experimental conditions. Using these approximations we show that the significant  properties of the measured Nernst effect are reproduced by the theoretical description which then justifies the simplifications made to solve the model. 

\begin{figure}
\includegraphics[width=\textwidth]{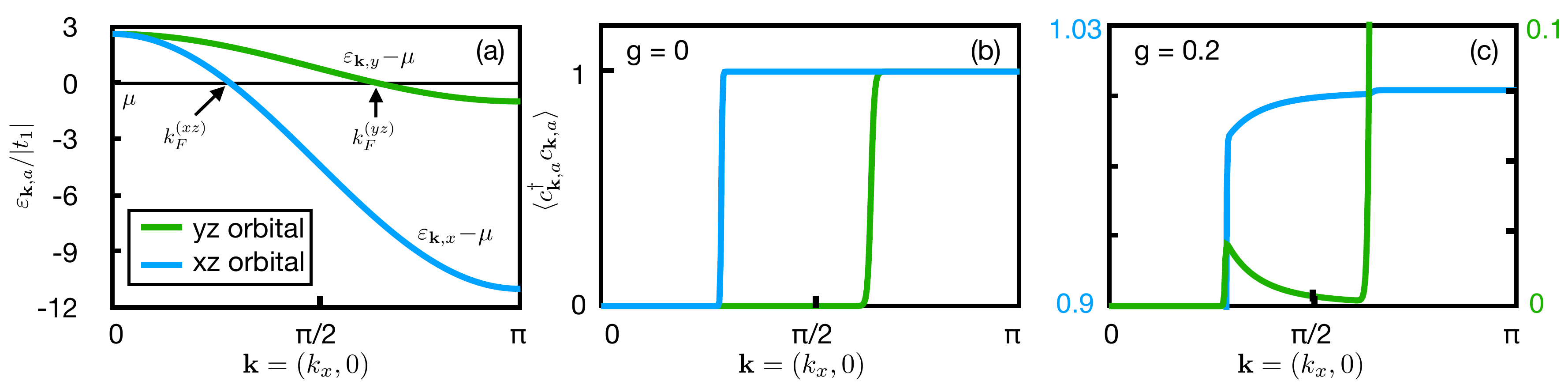}
\caption{Dispersion and occupation numbers of electrons in the $d_{xz}$ (blue solid lines) and in the $d_{yz}$ (green solid lines) orbital states plotted in momentum space along the $k_x$ direction. The chemical potential is fixed to $\mu = 1.2$. a) Difference between the orbital dispersions and the chemical potential, $\varepsilon_{{\bf k},a} - \mu$, calculated from Eqs.~\eqref{epsilonx},\eqref{epsilony}. Since $V_{\bf k}^t$ vanishes along the $k_x$-axis the dispersions $\varepsilon_{{\bf k},a}$ correspond in the absence of nematic fluctuations, i.~e.~$g=0$, to two bands leading to the well-known hole- and electron-like Fermi surface sheets around $(0,0)$ and $(\pi,0)$.  b) Orbital occupation numbers (per spin direction) as a function of momentum along the $k_x$-axis in the $g=0$ case and a finite small temperature $k_BT = 0.02$. Due to the absence of interactions these expectation values correspond to Fermi functions. c) Same quantities for $g = 0.2$. To highlight their momentum variations due to the nematic fluctuations the occupation numbers are plotted between 0.9 and 1.03 for the $d_{xz}$ orbital and between 0 and 0.1 for the $d_{yz}$ orbital.}
\label{Fig:Dispersion_1}
\end{figure}

The first step is to decouple the interaction term $\mathcal{H}_g$ of the model Hamiltonian \eqref{Model_H} using mean-field theory. After a Fourier transformation to the momentum space we obtain at first the following expression for the kinetic energy part of the Hamiltonian,
\begin{eqnarray}
\label{H_0}
\mathcal{H}_0 &=& \sum_{{\bf k},\sigma} \left[ (\varepsilon_{{\bf k},x} - \mu) c_{{\bf k}x\sigma}^\dagger c_{{\bf k}x\sigma}^{} +  (\varepsilon_{{\bf k},y} - \mu) c_{{\bf k}y\sigma}^\dagger c_{{\bf k}y\sigma}^{} + V_{\bf k}^t (c_{{\bf k}x\sigma}^\dagger c_{{\bf k}y\sigma}^{} + c_{{\bf k}y\sigma}^\dagger c_{{\bf k}x\sigma}^{}) \right],
\end{eqnarray}
where 
\begin{eqnarray}
\varepsilon_{{\bf k},x} &=& 2t_1 \cos k_x - 2t_2 \cos k_y + 2t_3 \left[ \cos (k_x + k_y) + \cos (k_x - k_y) \right] \label{epsilonx} \\
\varepsilon_{{\bf k},y} &=& 2t_1 \cos k_y - 2t_2 \cos k_x + 2t_3 \left[ \cos (k_x + k_y) + \cos (k_x - k_y) \right] \label{epsilony} \\
V_{\bf k}^t &=& 2t_4  \left[ -\cos (k_x + k_y) + \cos (k_x - k_y) \right]. \label{V_kt}
\end{eqnarray}
The hopping matrix elements $t_{1\dots 4}$ are fixed according to  Ref.~\cite{Yamase2013} to the following ratios with respect to $t_1$, $t_2 / t_1 = 1.5$, $t_3 / t_1 = 1.2$, and $t_4 / t_1 = 0.95$. The dispersions $\varepsilon_{{\bf k},x}$ and $\varepsilon_{{\bf k},y}$ are shown in Fig.~\ref{Fig:Dispersion_1}(a) for a specific momentum cut. 

From now on any energetic quantity will be given in units of $t_1$. The mean-field decoupled interaction term of Hamiltonian \eqref{Model_H} has the same operator structure as $\mathcal{H}_0$ and reads in mean-field approximation,
\begin{eqnarray}
\label{H_g}
\mathcal{H}_{g,MF} &\approx& g \sum_{{\bf k},\sigma} \bigg[ \left( n_x - \frac{1}{2} \right) c_{{\bf k}x\sigma}^\dagger c_{{\bf k}x\sigma}^{} + \left( n_y - \frac{1}{2} \right) c_{{\bf k}y\sigma}^\dagger c_{{\bf k}y\sigma}^{} \nonumber \\ 
&-& d_{xy} (c_{{\bf k}x\sigma}^\dagger c_{{\bf k}y\sigma}^{} + c_{{\bf k}y\sigma}^\dagger c_{{\bf k}x\sigma}^{}) \bigg] - gN (n_x^2 + n_y^2 - 2 n_{h}^2),
\end{eqnarray}
where $N$ is the total number of ${\bf k}$ points in the Brillouin zone and $n_x = \langle c_{{\bf i}x\sigma}^\dag c_{{\bf i}x\sigma}^{}\rangle$, $n_y = \langle c_{{\bf i}y\sigma}^\dag c_{{\bf i}y\sigma}^{}\rangle$, and $n_{h} = \langle c_{{\bf i}x\sigma}^\dag c_{{\bf i}y\sigma}^{}\rangle$ are expectation values of the local orbital occupation numbers and the local hybridization, respectively. These expectation values are formed with the approximated Hamiltonian $\mathcal{H}_{MF} = \mathcal{H}_0 + \mathcal{H}_{g,MF}$ and they have constant values for different lattice sites ${\bf i}$ and spin $\sigma$ due to the translation and time reversal symmetry of the original Hamiltonian \eqref{Model_H}. Spontaneous symmetry breaking of the mean-field Hamiltonian $\mathcal{H}_{MF}$ is not considered here since RPA predicts orbital order only for larger values of $g$.  

Considering the transport tensor coefficients in Eq.~\eqref{Model_H} within a semiclassical Boltzmann dynamics in the relaxation time approximation we obtain the following expressions ~\cite{Sharma2016},
\begin{eqnarray}
\sigma_{xx} &=& \frac{e^2\tau}{N k_B T} \sum_{{\bf k},a,\sigma} v_{xa}^2 n_{{\bf k}a} (1 - n_{{\bf k}a}) , \label{sigmaxx} \\
\alpha_{xx} &=& \frac{e\tau}{N (k_B T)^2} \sum_{{\bf k},a,\sigma} v_{xa}^2 (\varepsilon_{{\bf k},a} - \mu) n_{{\bf k}a} (1 - n_{{\bf k}a}), \label{alphaxx} \\
\sigma_{xy} &=& \frac{e^3\tau^2B}{\hbar N k_BT} \sum_{{\bf k},a,\sigma} \left(v_{xa}^2  \frac{\partial^2 \varepsilon_{{\bf k},a}}{\partial k_y^2} - v_{xa} v_{ya} \frac{\partial^2 \varepsilon_{{\bf k},a}}{\partial k_x \partial k_y} \right) n_{{\bf k}a} (1 - n_{{\bf k}a}) , \label{sigmaxy} \\
\alpha_{xy} &=& \frac{e^3\tau^2B}{\hbar N(k_B T)^2} \sum_{{\bf k},a,\sigma} \left(v_{xa}^2  \frac{\partial^2 \varepsilon_{{\bf k},a}}{\partial k_y^2} - v_{xa} v_{ya} \frac{\partial^2 \varepsilon_{{\bf k},a}}{\partial k_x \partial k_y} \right) (\varepsilon_{{\bf k},a} - \mu) n_{{\bf k}a} (1 - n_{{\bf k}a}),  \label{alphaxy}
\end{eqnarray}
where $\tau$ is the scattering time, $e$ the elementary charge, and $n_{{\bf k}a} =  \langle c_{{\bf k}a\sigma}^\dag c_{{\bf k}a\sigma}^{}\rangle$ the momentum-dependent occupation numbers of orbital '$a$'. The numbers $v_{xa} = \partial \varepsilon_{{\bf k},a} / \partial k_x$ and $v_{ya} = \partial \varepsilon_{{\bf k},a} / \partial k_y$ denote the components of the Fermi velocity with respect to orbital 'a'. Eqs.~\eqref{sigmaxx}-\eqref{alphaxy} have to be evaluated explicitly in order to calculate the Nernst coefficient $\nu$. Thereby, analytical expressions for the momentum derivatives of the dispersions $\varepsilon_{{\bf k},a}$ can be derived directly from Eqs.~\eqref{epsilonx}-\eqref{epsilony}. Here, only the hopping parameters $t_{1\dots4}$ enter. Thus, the only property which is determined by nematic fluctuations (represented by the parameter $g$) is the momentum-dependence of the occupation numbers $n_{{\bf k}a}$.

Expressions for the occupation numbers $n_{{\bf k}a}$ can be found easily by diagonalizing the mean-field Hamiltonian $\mathcal{H}_{MF}$. In the following, we quickly discuss the corresponding approach. Due to the bilinear operator structure of the parts $\mathcal{H}_0$ and $\mathcal{H}_{g,MF}$ the model Hamiltonian can be at first written in the form
\begin{eqnarray}
\mathcal{H}_{MF} &=& \sum_{{\bf k},a,\sigma} E_{{\bf k}a} c_{{\bf k}a\sigma}^\dagger c_{{\bf k}a\sigma}^{} + \sum_{{\bf k},\sigma} V_{\bf k} \left( c_{{\bf k}x\sigma}^\dagger c_{{\bf k}y\sigma}^{} + c_{{\bf k}y\sigma}^\dagger c_{{\bf k}x\sigma}^{} \right) + E_0,
\end{eqnarray}
where by use of Eqs.~\eqref{H_0} and \eqref{H_g} we have introduced
\begin{eqnarray}
E_{{\bf k}a} &=& \varepsilon_{{\bf k}a} + g \left( n_a - \frac{1}{2} \right) - \mu, \label{E_ka} \\
V_{\bf k} &=& V_{\bf k}^t - g n_{h}, \label{V_k} \\
E_0 &=& -gN \left( n_x^2 + n_y^2 - 2n_{h}^2 \right).
\end{eqnarray}
A diagonalization of $\mathcal{H}_{MF}$ can be achieved by using  the new fermionic operators 
\begin{eqnarray}
\alpha_{{\bf k}\sigma}^\dag &=& u_{\bf k} c_{{\bf k}y\sigma}^\dagger + v_{\bf k} c_{{\bf k}x\sigma}^\dagger \label{alpha}\\
\beta_{{\bf k}\sigma}^\dag &=& -v_{\bf k} c_{{\bf k}y\sigma}^\dagger + u_{\bf k} c_{{\bf k}x\sigma}^\dagger, \label{beta}\
\end{eqnarray}
where the momentum-dependent coefficients $u_{\bf k}$ and $v_{\bf k}$ have to fulfill the following properties,
\begin{eqnarray}
u_{\bf k}^2 &=& \frac{1}{2} \left( 1 - \frac{E_{{\bf k}x} - E_{{\bf k}y}}{\sqrt{(E_{{\bf k}x} - E_{{\bf k}y})^2 + 4 V_{\bf k}^2}} \right) \label{coeff_u}\\
v_{\bf k}^2 &=& 1 - u_{\bf k}^2 \label{coeff_v}
\end{eqnarray}
Under these conditions we find for the mean-field Hamiltonian a diagonal operator structure,
\begin{eqnarray}
\label{H_MF}
\mathcal{H}_{MF} &=& \sum_{{\bf k},\sigma} \left(E_{\bf k}^{(e)} \alpha_{{\bf k}\sigma}^\dag \alpha_{{\bf k}\sigma}^{} + E_{\bf k}^{(h)} \beta_{{\bf k}\sigma}^\dag \beta_{{\bf k}\sigma}^{} \right) + E_0,
\end{eqnarray}
and the Hamiltonian is now written in the 'band' basis where '$e$' denotes the electron pocket and '$h$' the hole pocket in the conventional notation of the iron-based superconductors. The quasiparticle dispersions read,
\begin{eqnarray}
\label{E_k^eh}
E_{\bf k}^{(e,h)} &=& \frac{E_{{\bf k}x} + E_{{\bf k}y}}{2} \pm \sqrt{(E_{{\bf k}x} - E_{{\bf k}y})^2 + 4 V_{\bf k}^2},
\end{eqnarray} 
where the dispersion functions $E_{\bf k}^{(e,h)}$ now describe the actual band structure (for example measurable by ARPES) consisting of two bands with the typical Fermi surface topology in iron-based superconductors. Explicit expressions for the momentum-dependence of $E_{\bf k}^{(e,h)}$ can be derived from Eq.~\eqref{E_k^eh} by use of the Eqs.~\eqref{E_ka},\eqref{V_k} and Eqs.~\eqref{epsilonx}-\eqref{V_kt}. In Fig.~\ref{Fig:Dispersion_1}(a) the band structure is plotted for the $g=0$ case.

We now turn to the actual calculation of the expectation values $\langle c_{{\bf k}a}^\dag  c_{{\bf k}a}^{} \rangle$ and $\langle c_{{\bf k}x}^\dag  c_{{\bf k}y}^{} \rangle$ which enter all transport coefficients in Eqs.~\eqref{sigmaxx}-\eqref{alphaxy} and also the local expectation values $n_a = (1/N) \sum_{\bf k} \langle c_{{\bf k}a}^\dag  c_{{\bf k}a}^{} \rangle$ and $n_h = (1/N) \sum_{\bf k} \langle c_{{\bf k}x}^\dag  c_{{\bf k}y}^{} \rangle$. Since the Hamiltonian $\mathcal{H}_{MF}$ is diagonal in the formulation \eqref{H_MF} the above expectation values can be easily calculated when the $c$-operators are replaced by the quasiparticle operators $\alpha$ and $\beta$ using Eqs.~\eqref{alpha} and \eqref{beta}. The remaining expectation values are bilinear in the quasiparticle operators and reduce to simple Fermi functions with respect to the dispersions in Eq.~\eqref{E_k^eh}. We obtain
\begin{eqnarray}
\langle c_{{\bf k}x}^\dag  c_{{\bf k}x}^{} \rangle &=& u_{\bf k}^2 \, f(E_{\bf k}^{(h)}) + v_{\bf k}^2 \, f(E_{\bf k}^{(e)}), \label{nx} \\
\langle c_{{\bf k}y}^\dag  c_{{\bf k}y}^{} \rangle &=& v_{\bf k}^2 \, f(E_{\bf k}^{(h)}) + u_{\bf k}^2 \, f(E_{\bf k}^{(e)}), \label{ny} \\
\langle c_{{\bf k}x}^\dag  c_{{\bf k}y}^{} \rangle &=& u_{\bf k} v_{\bf k} \left[f(E_{\bf k}^{(e)}) -  f(E_{\bf k}^{(h)}) \right], \label{dxy}
\end{eqnarray}
where $f(E) = 1 / (1 + e^{\beta E})$ is the Fermi function and $u_{\bf k}, v_{\bf k}$ are given by the Eqs.~\eqref{coeff_u} and \eqref{coeff_v}.

\subsection{Numerical results}

We now show by numerical evaluation of the expectation values \eqref{nx}-\eqref{dxy} that the Nernst effect becomes dramatically larger in the presence of nematic fluctuations. For this we have calculated the transport coefficients \eqref{sigmaxx}-\eqref{alphaxy} and the Nernst coefficient $\nu$ for given sets of the parameters $\mu$, $g$, and $T$. 

\begin{figure}
\includegraphics[width=0.4\textwidth]{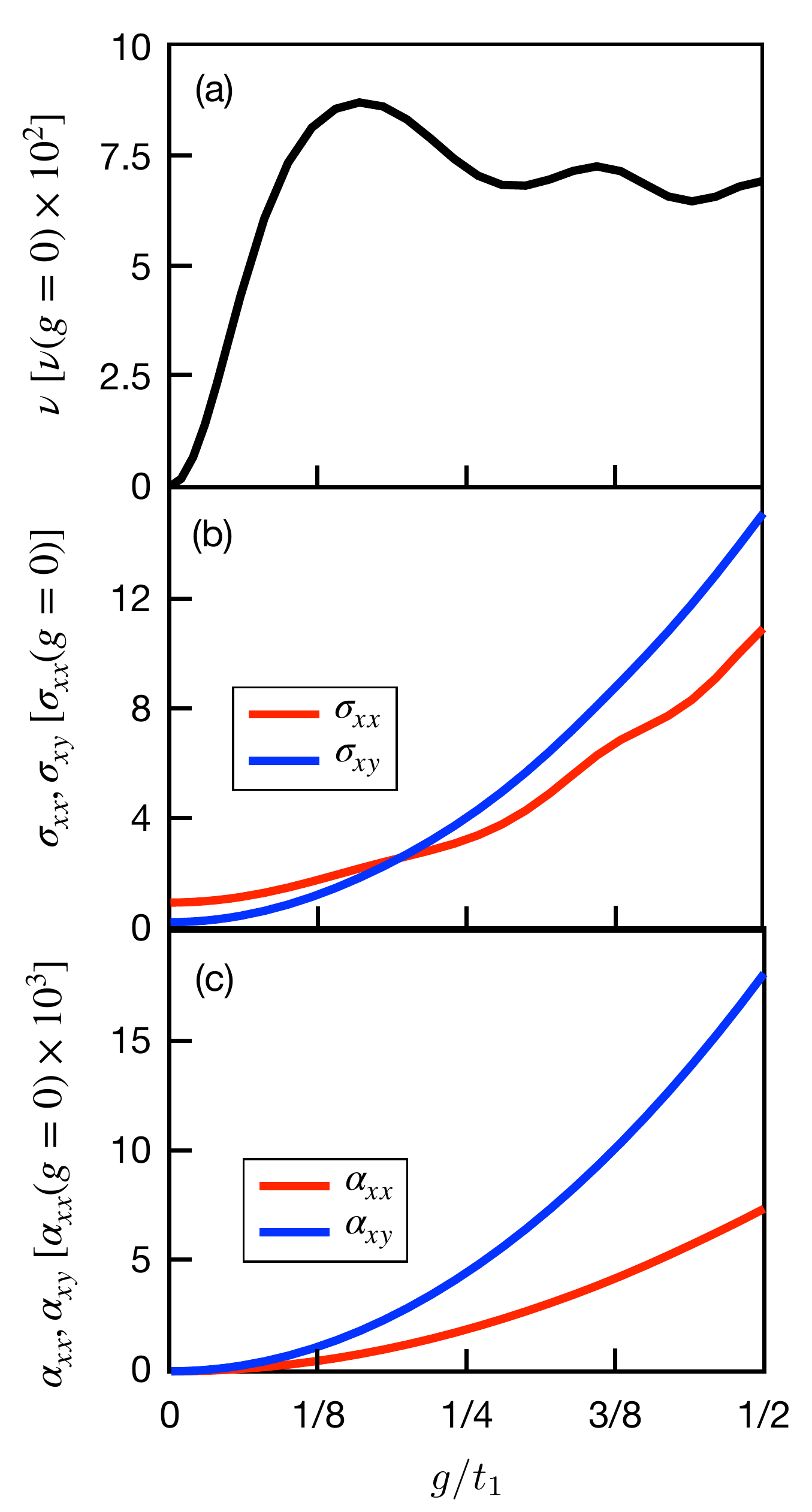}
\caption{Calculated Nernst coefficient and transport components as a function of the strength $g$ of the nematic fluctuations for $\mu = 1$ and $k_BT = 0.01$. a) Calculated Nernst coefficient $\nu$. Starting from very small values $\nu$ increases dramatically already for relatively small values of $g$. b) Electrical conductivity $\sigma_{xx}$ (red line) and Hall conductivity $\sigma_{xy}$ (blue line) as a function of $g$. The Hall coefficient grows more rapidly than $\sigma_{xx}$ and eventually exceeds its absolute value at $g/t_1 \approx 0.2$. This causes for $\nu$ a change from a strong increase followed by a saturation. c) Peltier coefficients $\alpha_{xx}$ (red line) and $\alpha_{xy}$ (blue line) as a function of $g$. Also here, the anisotropic component is dominant in the presence of nematic fluctuations.}
\label{Fig:Nernst_effect}
\end{figure}

We start with a discussion how $\nu$ generally behaves when the strength $g$ of the nematic fluctuations is slowly switching from $g=0$ to finite but still relatively small values. Fig.~\ref{Fig:Nernst_effect}(a) shows the calculated $\nu$ as a function of $g$ for a fixed value of the chemical potential, $\mu = 1$, and a fixed small temperature value of $k_BT = 0.01$. One can clearly see that $\nu$ increases strongly already in a relatively narrow $g$ range, $0 < g < 0.2$, before it reaches a saturation. We particularly note that already at $g = 1/8$ the absolute value of $\nu$ becomes three orders of magnitude larger than its initial value in the interaction-free case. In Figs.~\ref{Fig:Nernst_effect}(b,c) the transport coefficients which determine $\nu$ via Eq.~\eqref{nu_general} are plotted separately. It can be seen that both the Hall conductivity $\sigma_{xy}$ and also the anisotropic Peltier component $\alpha_{xy}$ (blue lines) grow rapidly with $g$. Their increase is even stronger than that of the conventional longitudinal  transport components $\sigma_{xx}$ and $\alpha_{xx}$ (red lines). In the zero-$g$ limit the Hall conductivity is much smaller than $\sigma_{xx}$ but due to the stronger increase it exceeds $\sigma_{xx}$ for $g$-values larger than $g \approx 0.2$. This specific behavior of the transport coefficients influences the Nernst coefficient $\nu$ significantly since it is determined by the ratio~\eqref{nu_general} where a summation of different tensor components enter in different ways in the nominator and denominator. It leads for $\nu$ to a change from strong increasing to rather saturating behavior appearing around $g \approx 0.2$ where $\sigma_{xx}$ and $\sigma_{xy}$ coincide.  

The strong impact of nematic fluctuations to the anisotropic transport coefficients can be understood as follows. A closer inspection of the Eqs.~\eqref{sigmaxy},\eqref{alphaxy} shows that the lowest-order effect of nematic fluctuations is determined by the momentum-dependent occupation numbers $n_{{\bf k}a}$. All other quantities including  dispersions $\varepsilon_{{\bf k},a}$ and their derivatives are independent of $g$. In Fig.~\ref{Fig:Dispersion_1}(b,c) we show the momentum-dependence of the $n_{{\bf k}a}$ for $g=0$ (panel (b)) and $g=0.2$ (panel (c)) where the Nernst coefficient has already significantly increased according to Fig.~\ref{Fig:Nernst_effect}(a). In the interaction-free case the occupation numbers are just the Fermi distribution functions separating fully occupied states (at $T=0$), defined by the property $\varepsilon_{{\bf k},a} < \mu$, from the unoccupied states. The factor $n_{{\bf k}a} (1 - n_{{\bf k}a})$ in the momentum summation in Eqs.~\eqref{sigmaxy},\eqref{alphaxy} is non-zero only in a narrow ${\bf k}$-range around the corresponding Fermi momentum $k_F^{(a)}$ (compare panel (a)) where $\varepsilon_{{\bf k},a} - \mu$ is of the order of the thermal broadening $k_BT$. This limitation of the contributing terms in the momentum summation is well-known and typical for the conventional thermal and electrical transport of conduction electrons in usual metals. It is for example the reason why the Nernst effect is usually very small.  

Switching on the nematic fluctuations $g$ changes  the situation dramatically as is seen in panel (c). Due to the coupling between the two different types of orbitals there is now an increase of the $d_{yz}$ occupation (green solid line) around the Fermi momentum $k_F^{(xz)}$ of the {\it other} orbital, i.~e. the $d_{xz}$ orbital. Note that the $d_{yz}$ dispersion  $\varepsilon_{{\bf k},yz}$ is far above the Fermi level in this momentum region and thus would not be occupied in the decoupled case at low temperature. This non-zero occupation of $d_{yz}$, which is solely due to the nematic coupling proportional to $g$, is relatively small but it is not bounded to some region around the Fermi level. Instead it affects the whole momentum range where the dispersion of the other orbital $d_{xz}$ is below the Fermi level. This leads to a macroscopic number of ${\bf k}$-points where $n_{{\bf k}a} (1 - n_{{\bf k}a})$ is non-zero and thus the number of contributing terms in the momentum summations~\eqref{sigmaxy},\eqref{alphaxy} increases dramatically. Furthermore, as a consequence of the occupation $n_{{\bf k}yz}$, the number of $d_{xz}$ electrons, $n_{{\bf k}xz}$, (blue solid line) in the same ${\bf k}$-state is reduced by  $n_{{\bf k}yz}$. Finally, the occupation of both orbitals changes within the thermal broadening to 1 when $k_F^{(yz)}$ is reached. Note that the same discussion holds for the perpendicular direction in the ${\bf k}$-space but with the opposite orbital characteristics. 

\begin{figure}
\includegraphics[width=\textwidth]{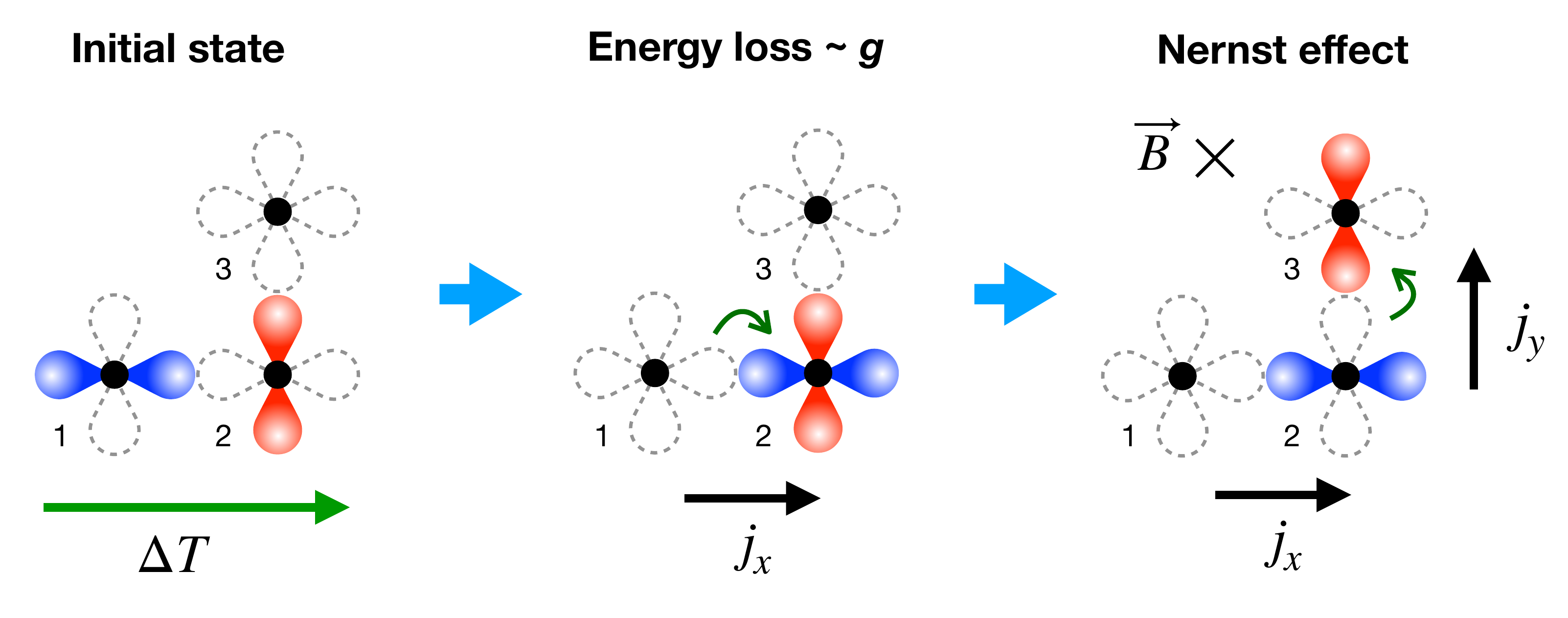}
\caption{Schematic picture of the enhanced Nernst effect as driven by nematic fluctuations. An external temperature gradient $\Delta T$ in $x$-direction generates  a preferred hopping of  conduction electrons  in the  $d_{xz}$ orbital (highlighted by blue color). One of such electrons sits for example at some lattice site '1' (left panel). If the neighboring lattice site '2' is already  occupied with a $d_{yz}$ orbital (highlighted by red color) an intermediate state with double occupancy including both orbital types is created (middle panel) after hopping from '1' to '2'. Due to the nematic interaction in Eq.~\eqref{Model_H}  this state has an of the order of $g$ higher energy than the initial state. The enhanced energy can be reduced again by either further hopping in $x$-direction or (most importantly) removing the electron in the $d_{yz}$ orbital from site '2'. Due to its orbital character this electron preferably hops in $y$-direction to site '3' where the hopping direction is pretended by the Lorentz force of the perpendicularly oriented magnetic field $\vec{B}$ (right panel). The preferred transport in $y$-direction causes an enhanced Nernst effect.}
\label{Fig:Picture}
\end{figure}

In conclusion, the nematic coupling leads in the momentum range where for zero-coupling one orbital is occupied and the other one is empty to a fluctuation-driven 'flow' from the occupied to the unoccupied orbital which is oriented towards the transverse  direction. Since this correlation affects also higher energy states it has a dramatic effect on the anisotropic transport coefficients which particularly describe the response to a flow in the perpendicular direction. Fig.~\eqref{Fig:Picture} shows the schematic picture of the responsible processes leading to this effect. An external field (for example temperature gradient) applied in $x$-direction generates at first a  hopping of  conduction electrons  in $x$ direction. This can lead to an intermediate state with double occupancy including both orbital types. Due to the nematic interaction in Eq.~\eqref{Model_H}  this state has an of the order of $g$ higher energy than the initial state. The enhanced energy can be reduced again by either further hopping in $x$-direction or (most importantly) removing the electron in the $d_{yz}$ orbital from the doubly occupied site. Due to its orbital character this electron preferably hops in $y$-direction where the hopping direction is pretended by the Lorentz force of the perpendicularly oriented magnetic field. The resulting preferred transport in $y$ direction explains the high sensitivity of the Nernst effect towards nematic fluctuations. 

\begin{figure}
\includegraphics[width=\textwidth]{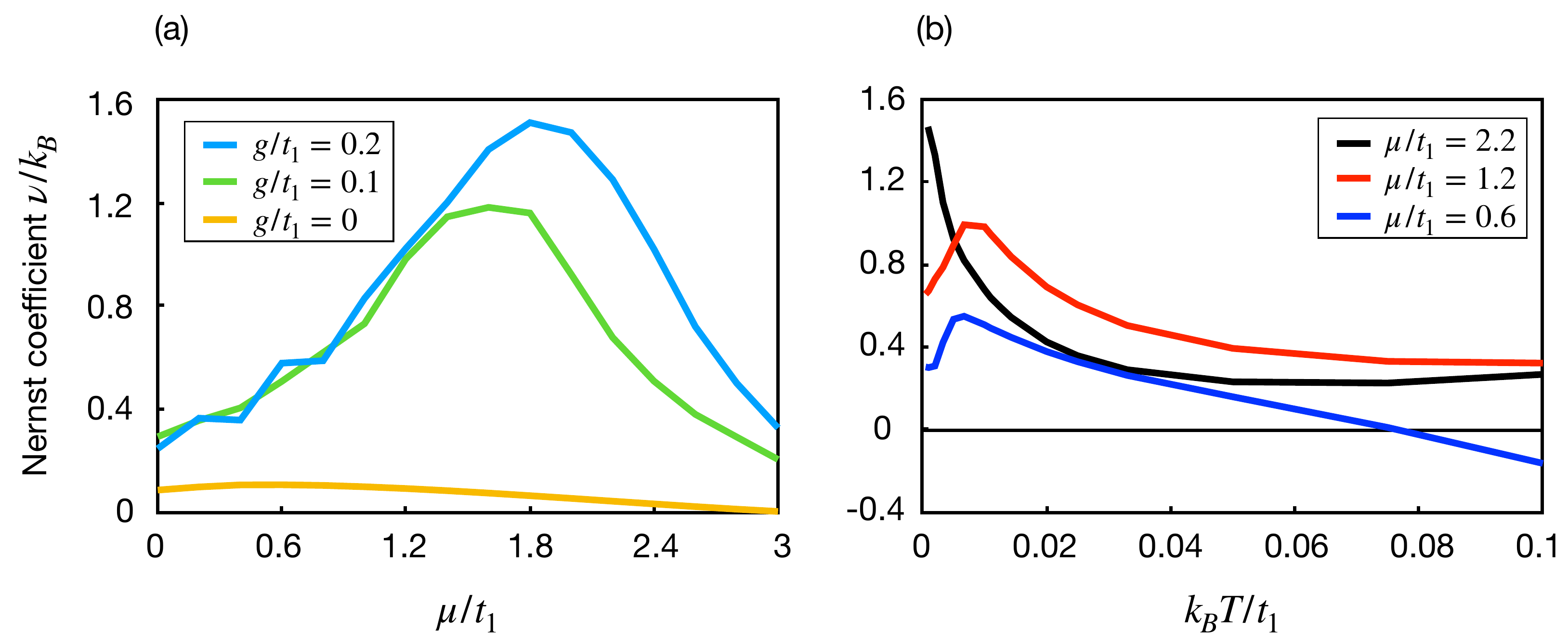}
\caption{Calculated Nernst coefficient under variation of further parameters. a) Nernst coefficient as a function of the chemical potential $\mu$ for two different values of $g$ and fixed temperature $k_BT = 0.01$. The broad maximum at intermediate $\mu$ values (for $g\neq 0$) is comparable with the observed behaviour of the Nernst effect as a function of doping as it has been found by our measurements. For comparison we also show the result without taking into account nematic coupling ($g=0$), which shows a very reduced amplitude. b) Nernst coefficient as a function of temperature for different values of $\mu$ and fixed $g=0.1$. As observed in the experiments we find a $\mu$-dependent low temperature maximum followed by a rapid decrease and saturation at high $T$. The maximum appears when the thermal energy $k_BT$ is of the same order as the experimentally found value $k_BT_c$ for the superconducting $T_c$.}
\label{Fig:Variation_mu_T}
\end{figure}

Finally, in Fig.~\ref{Fig:Variation_mu_T} we show the calculated Nernst coefficient $\nu$ as a function of the two further free parameters, the chemical potential $\mu$ and temperature $T$. One can clearly see from Fig.~\ref{Fig:Variation_mu_T}(a) that $\nu$ is mostly enhanced in a rather broad range around $\mu \approx 1.8$. Since $\mu$ is closely related to the electron filling its variation can be considered as a simulation of doping the material with electrons or holes. Having this relationship in mind we argue that the calculated maximum of $\nu$ is consistent with the measured behavior of the Nernst effect for samples with different doping. The calculated influence of temperature  is shown in Fig.~\ref{Fig:Variation_mu_T}(b) for three different $\mu$-values. This result is also in qualitative agreement with our experiments. We find a $\mu$-dependent maximum at rather low temperatures $k_BT \approx 0.01$ which corresponds to the temperature scale of the superconducting critical temperatures measured for this material. At higher temperatures $\nu$ decreases again followed by a saturation or even negative values in the high-temperature limit.  

\begin{figure}
\includegraphics[width=0.5\textwidth]{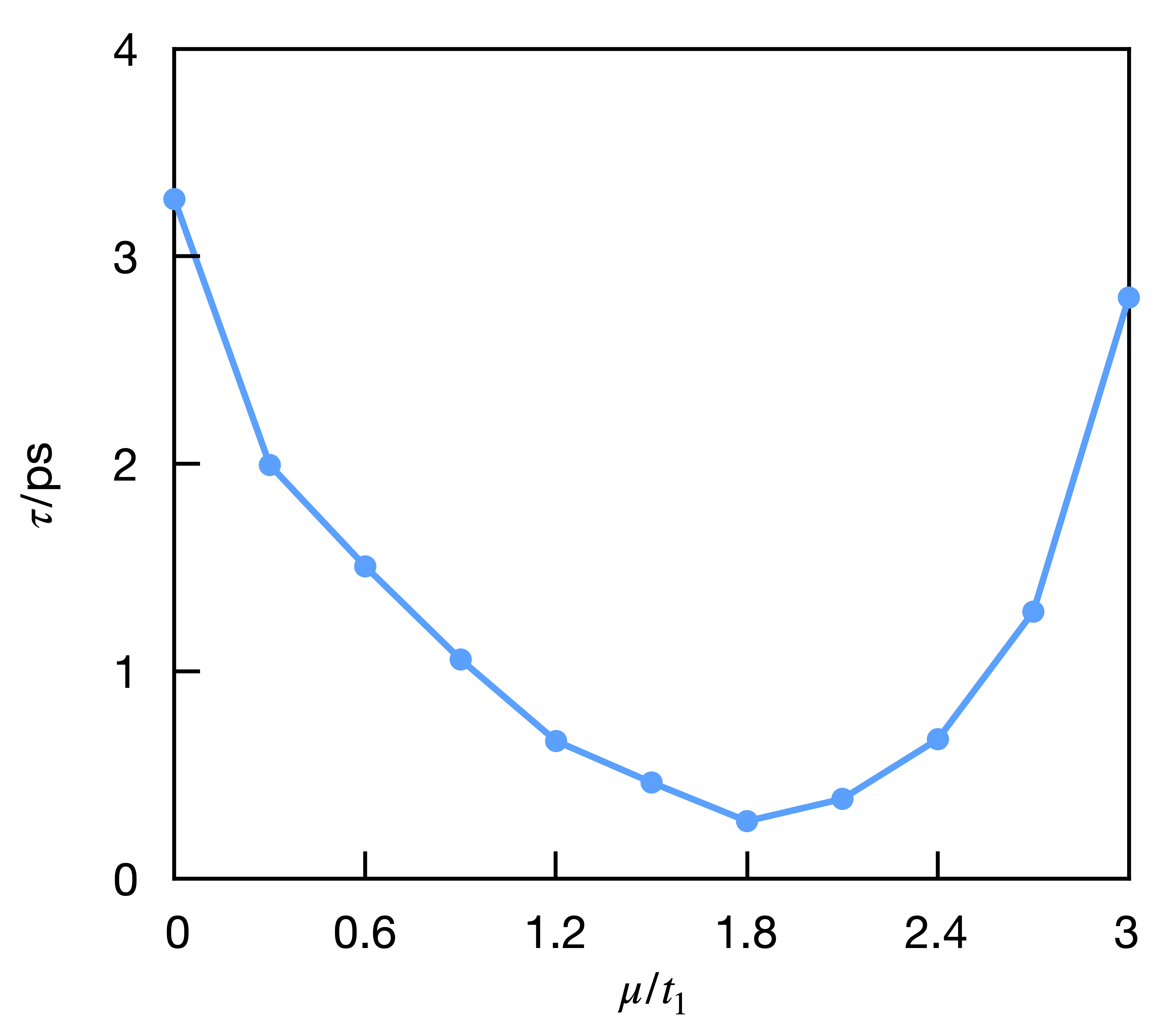}
\caption{Scattering time as a function of the chemical potential $\mu$ extracted from Eq.~\eqref{nu_general} being changed in such a way that $\nu$ remains constant. The coupling strength is fixed to $g/t_1 = 0.2$.}
\label{Fig:Scatt_time}
\end{figure}

Finally, to study the effect of the scattering rate to our observed effect we have artificially varied the scattering time $\tau$ in Eqs.~\eqref{sigmaxx}-\eqref{alphaxy} such that the calculated Nernst coefficient in Eq.~\eqref{nu_general} remains constant. The result is shown in Fig.~\ref{Fig:Scatt_time}. This corresponds to the change of $\nu$ which would be necessary to overcompensate the effect of the enhancement of the Nernst coefficient by nematic fluctuations. It is seen that the change of $\tau$ must be extremely large, i.e. about one order of magnitude.

\section{Crystal Growth}

Single crystals of \laco were grown via Na-As liquid assisted growth using diffusion controlled abnormal grain growth from polycrystalline \la by the solid state crystal growth method (SSCG) as described in Ref. \onlinecite{Kappenberger2018}. Single crystals of undoped \ba were grown by a high temperature solution growth using an excess of FeAs as solution (self-flux), (see, e.g., Ref. \onlinecite{Selter2020} and references therein). Single crystals of \barh were grown out of excess (Fe$_{1-x}$Rh$_x$)As using the procedures outlined in references \onlinecite{Canfield2019}, \onlinecite{Ni2008}, and \onlinecite{Ni2009}.

Crystals were analyzed using several complementary techniques as scanning electron microscopy (SEM) with energy dispersive x-ray analysis (EDX), Laue backscattering, powder and/or single crystal X-ray diffraction and SQUID magnetometry measurements.

\section{Nernst effect measurements}

\begin{figure}
\includegraphics[width=\textwidth]{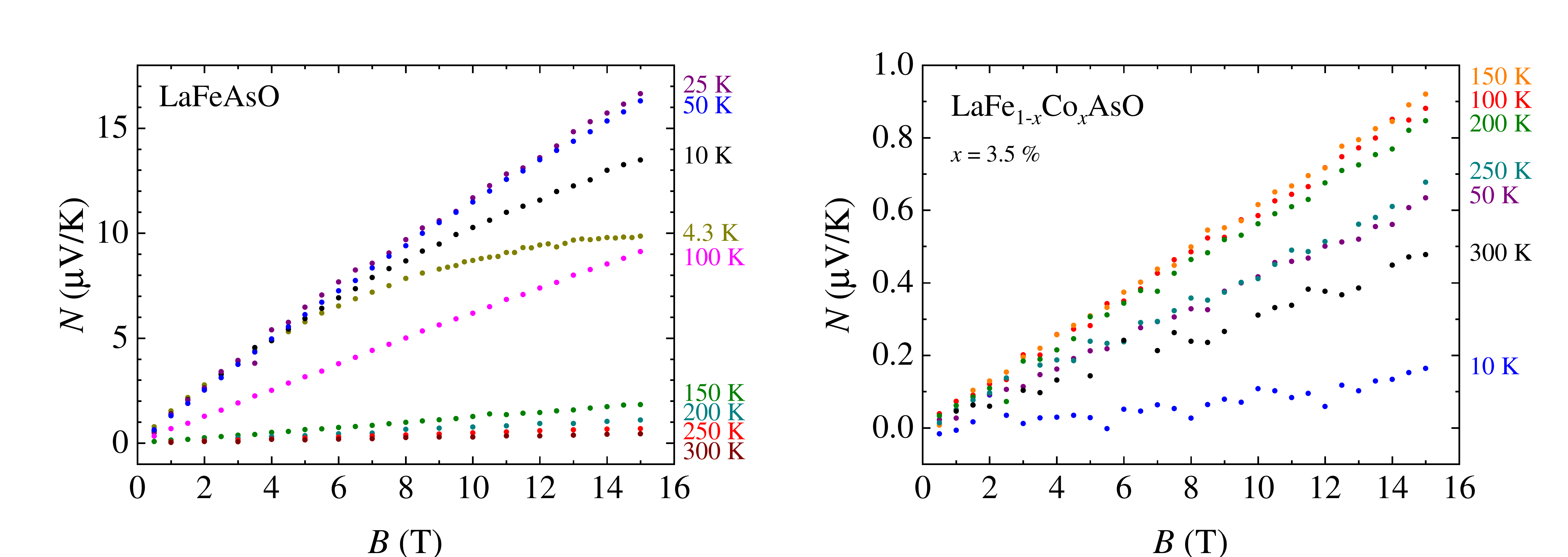}
\caption{Nernst signal as a function of magnetic field $B$ for \laco with \textbf{(a)} $x = 0$ and \textbf{(b)} $x = 3.5~\%$ at different constant temperatures.} 
\label{Fig:Nernst_B}
\end{figure}

The measurements were conducted in a home-built device in a helium bath cryostat under magnetic fields in the range $\pm 15$~T. The single crystals were prepared in an inert atmosphere to prevent degradation. The thermal gradient in the crystals has been applied parallel to the $ab$-plane, using a resistive chip heater. The magnetic field along the $c$-axis and the Nernst voltage was measured orthogonal to the two former directions but again in the $ab$-plane, using a Keithley nanovoltmeter. The measured voltage has then been antisymmetrized with respect to magnetic field. All measured samples show a linear behaviour of the Nernst signal as a function of field (compare Fig S5, which exemplary shows $N(B)$ for two different doping levels of \laco) in the tetragonal state, which is of particular interest for our analysis.

\section{Elastoresistivity measurements}

\begin{figure}
\includegraphics[width=\textwidth]{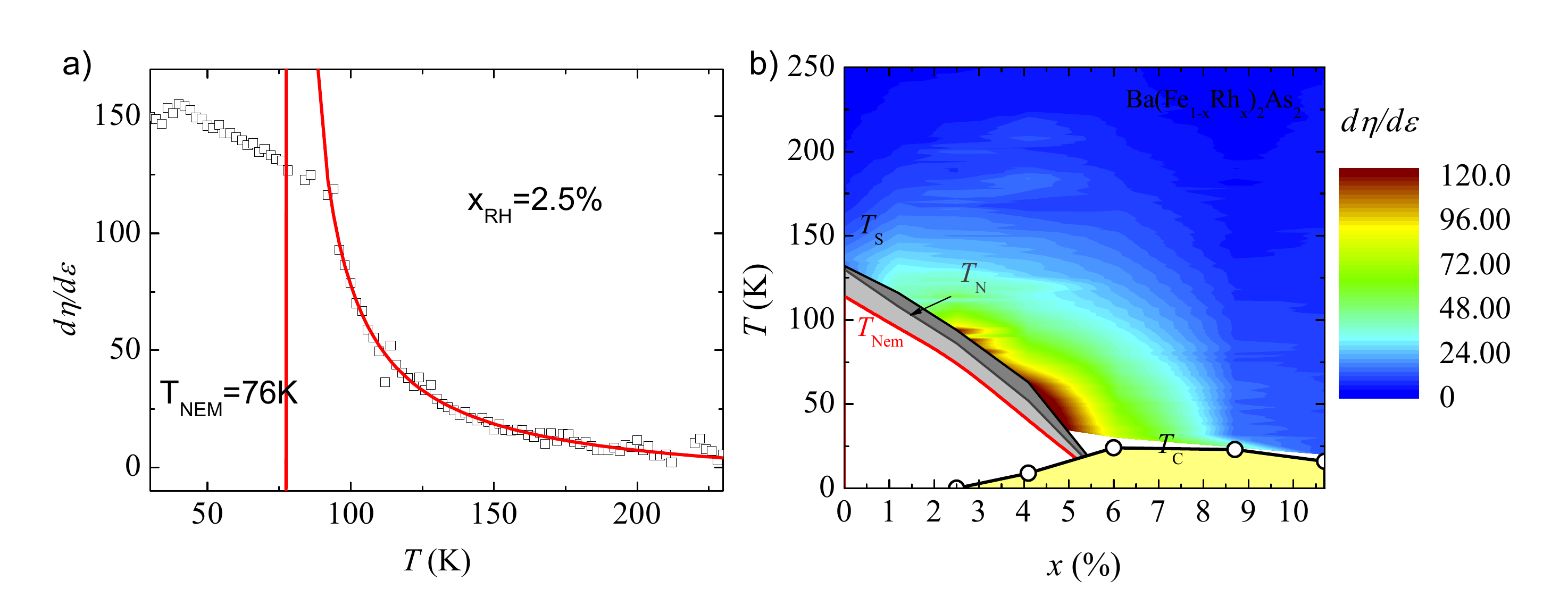}
\caption{Elastoresistance of the Ba(Fe$_{1-x}$Rh$_x$)$_2$As$_2$ series. \textbf{(a)} Temperature-dependence of the strain ($\epsilon$) derivative of the relative variation of the resistivity ($\eta$=$\Delta \rho$/$\rho$) for a sample with x=0.025. The red curves shows the Curie-Weiss fit in the nematic fluctuation regime above the structural transition. \textbf{(b)} Phase diagram of the nematic susceptibility of the Ba(Fe$_{1-x}$Rh$_x$)$_2$As$_2$ series. T$_C$, T$_N$, T$_S$ and T$_{Nem}$  represent the superconducting, the magnetic, the structural and the nematic critical temperature, respectively. The color scale reproduce the magnitude of $d\eta/d\epsilon$, while the light yellow region corresponds to the superconducting dome.} 
\label{Fig:ElastoR}
\end{figure}

We measured the elastoresitivity of all the samples of the Ba(Fe$_{1-x}$Rh$_x$)$_2$As$_2$ (with x=0, 0.012, 0.025, 0.04, 0.061, 0.087, 0.0107) series, in order to obtain the nematic phase diagram (Figure S6b) discussed in the main text. For the measurement we adopted the standard procedure described in Ref \onlinecite{Chu2012} and \onlinecite{Hong2020}. The samples have been glued to a commercial piezoelectric actuator, oriented in order to have the crystalline [110] direction aligned along the main straining-axis of the piezo stack. The applied strain has been measured with a strain gauge attached to the same piezo actuator. Figure S6a shows the temperature-dependence of the strain derivative of the resistivity $d\eta/d\epsilon$ measured on a sample with x=0.025, as an example. Every point has been obtained at fixed temperature, by scanning the voltage applied to the piezo actuator from -30 V to 100 V and detecting the corresponding variation in $\eta$ and $\epsilon$. Due to the tiny applied strain (ranging from 0.01$\%$ to 0.1$\%$) we have  operated in a linear regime of $\eta$ vs $\epsilon$. The nematic temperature T$_{Nem}$ has been obtained as a fitting parameter by interpolating the temperature-dependence of $d\eta/d\epsilon$ in the fluctuating regime with a Curie-Weiss function $d\eta/d\epsilon$=$C_0$+$C/(T-T_{Nem})$ (red curve in Figure S6a) ~\cite{Chu2012, Hong2020}.


%